\documentclass[10pt]{article}
\usepackage{amsmath}
\usepackage{breqn}
\usepackage{xcolor}
\usepackage{cite}

\usepackage{booktabs, array, makecell, bm}
\usepackage{titlesec}

\titleformat{\section}
  {\large\bfseries}   % Burada boyut + stil
  {\thesection}       % Numara
  {1em}               % Numara ile başlık arasındaki boşluk
  {}            
\newcolumntype{C}[1]{>{\centering\arraybackslash}m{#1}}
\begin{document}

\title{\Large \bf From MOND entropy to extended uncertainty principles: A unified framework}

\author{{\"{O}zg\"{u}r Sevinç$^{a}$ \thanks{Email: ozgur.sevinc@yeniyuzyil.edu.tr}\hspace{1mm},\"{O}zg\"{u}r \"{O}kc\"{u}$^{a}$ \thanks{Email: ozgur.okcu@yeniyuzyil.edu.tr (Corresponding author)}\hspace{1mm} Ekrem Aydiner$^{b,c}$ \thanks{Email: aydiner@princeton.edu}} \\ $^a${\small {\em Department of Electrical and Electronic Engineering, Istanbul Yeni Yuzyil University, }}\\{\small {\em İstanbul 34010, Türkiye }} \\ $^b${\small {\em Department of Physics, Princeton University, Princeton, }}\\{\small {\em NJ 08544, USA}}\\ $^c${\small {\em Department of Physics, Faculty of Science, İstanbul University, }}\\{\small {\em İstanbul 34134, Türkiye}} }

\date{}

\maketitle

\begin{abstract}
In this study, we explore the relation between generalised entropies and the extended uncertainty principle (EUP) models. Starting from the higher-order extended uncertainty principle (HOEUP), we obtain the modified entropy-area relation. Then, we derive the modified Friedmann equations through three different approaches: the first law of thermodynamics at the apparent horizon, the entropic gravity case, and the emergence of cosmic space.  Furthermore, we check the validity of the generalised second law (GSL). Notably, HOEUP modified Friedmann equations are the limiting cases of those obtained from a recently proposed novel entropy, which is derived from Modified Newtonian Dynamics (MOND) [{\it Phys. Dark Universe} {\bf 49} (2025) 101967]. Motivated by this connection, we derive a novel EUP, referred to as MOND EUP, from a reverse procedure.  This novel EUP reproduces to EUP relations associated with Rényi and dual Kaniadakis entropies in the limiting cases. Moreover, we show that HOEUP corresponds to perturbative limit of MOND entropy. The main new result of this paper is a reverse procedure beginning from a recently proposed novel MOND entropy to construct a unified EUP. This reverse procedure is not limited with the present case. In principle, the method can be applied to other generalised entropy formalisms, suggesting that  our findings may establish a unified framework that bridges the generalised entropies, cutoff mechanisms, and EUP models. In particular, the corresponding modified uncertainty principles may have  effective cutoff mechanisms for the entropy forms, which  do not explicitly display cutoff mechanisms. Thus, these entropies may have cutoff mechanism due to their corresponding modified uncertainty principles.
\end{abstract}

\section{Introduction}
\label{intro}

Although the standard uncertainty principle successfully describes quantum systems at the atomic and subatomic scales, it faces failure when applied to the Planck scale or the cosmological scale. In these extreme scales, the quantum gravitational effects are expected to bring significant corrections to the standard uncertainty principle.  These corrections basically emerge as modifications of the momentum and position uncertainties. The first is usually referred to the generalised uncertainty principle (GUP) and the latter is usually referred to EUP. A more complete framework is referred to the generalised and extended uncertainty principle (GEUP) when both modifications are taken into consideration. A natural cutoff mechanism both at Planck scale or cosmological scale may emerge as a result of modified uncertainty principle. The GUP models usually imply the ultraviolet (UV) cutoff, which yields minimum length scale (typically Planck length) or equivalently maximum momentum scale. On the other hand, the EUP models usually imply infrared (IR) cutoff, which corresponds to large distance scales of the system (typically cosmological length),   that is, the existence of maximum length scale or equivalently minimum momentum scale. These cutoff mechanisms may play an important role both in black hole thermodynamics \cite{Adler2001,Nouicer2006,Sakalli2016,Xiang2009,Okcu2020,Scardigli2020a,Hassanabadi2019,Mureika2019,Okcu2022} and in the evolution of the Universe \cite{Awad2014,Salah2017,Okcu2020b,Alsabbagh2023,Okcu2024,Luo2023,Kouwn2018,Das2022BG,Feng2022,Luciano2021,Luo2025,Luciano2025,Scardigli2011,Perivolaropoulos2017}. For example, one of the most characteristic features emerges as black hole remnants at the final stage of evaporation when GUP effects are taken into account \cite{Adler2001}. On the other hand, a minimum temperature of black hole exists in the EUP case \cite{Hassanabadi2019}. EUP also affects black hole solutions and its characteristic properties such as  horizon radius, ISCO, and photosphere \cite{Mureika2019}. Moreover, within the framework of EUP, black holes in flat spacetime exhibit  phase transitions similar to those observed for AdS black holes \cite{Okcu2022}. 

Apart from black hole thermodynamics, the cosmological applications of the modified uncertainty principles have been extensively investigated in the literature. In particular, GUP models provide effective ways to cure the initial singularity \cite{Awad2014,Salah2017,Okcu2020b,Alsabbagh2023,Okcu2024}. Moreover, gravitational baryogenesis \cite{Das2022BG,Feng2022,Luo2023} and Big Bang Nucleosynthesis \cite{Luciano2021,Luo2025} have also been investigated within the framework of GUP and EUP, with the aim of revealing possible imprints of quantum gravity corrections on baryon asymmetry and primordial element abundances. In this context, the authors of Ref. \cite{Luciano2025} showed that GUP exhibits an effective quintessence-like dark energy behaviour and affects the growth of matter perturbation. In Ref. \cite{Scardigli2011}, the authors showed that the micro black hole remnants, which originate from GUP, may dominated the Universe before the beginning of inflation, leading to a pre-inflationary matter era \footnote{These studies show that various modifications of uncertainty principle play an important role both in black hole thermodynamics and cosmology. Various modifications of uncertainty principle have been investigated in the literature \cite{Perivolaropoulos2017,Maggiore1993,Scardigli1999,Kempf1995,Das2008,Bambi2008,Dabrowski2019,Du2022}.}.

While the modifications of uncertainty principle incorporate the quantum effects into gravitation, another way to combining gravitation with quantum mechanics is gravity-thermodynamics conjecture, which is fundamentally based on black hole thermodynamics \cite{Bekenstein1973,Hawking1975}. This conjecture generally refers to approaches that aim to derive the Friedmann equations, field equations, etc. from the underlying connection between gravity and thermodynamics. The research in this direction was pioneered by Jacobson \cite{Jacobson1995}, who showed that the Einstein field equation can be derived from the entropy-area relation $S=A_{h}/4$ and  Clausius relation $\delta Q=T_{h}dS_{h}$. Here, $\delta Q$ and $T_{h}$ are energy flux and Unruh temperature, respectively. Since then, gravity-thermodynamics conjecture has been an active field of research \cite{Cai2005,Akbar2007,Saridakis2020,Sheykhi2022,Saridakis2021,Sheykhi2021,Sheykhi2018,Lymperis2018,Sheykhi2023b,Lymperis2021,Almada2022,Almada2022b,Abreu2022,Odintsov2023,Coker2023,Okcu2024b,Ourabah2024,Verlinde2011,Cai2010,Shu2010,Sheykhi2010c,Sheykhi2011,Gao2010,Ling2010,FCai2010,Basilakos2012,Senay2021,Feng2018,Abreu2018,Jusufi2023,Padmanabhan2012,Sheykhi2013,Cai2012,Yang2012,Eune2013,Dezaki2015}. Motivated by Jacobson's idea, Cai et al. obtained the Friedmann equations from the first law of thermodynamics at the apparent horizon \cite{Cai2005,Akbar2007}. Subsequently, their approach was generalised to various modified gravity theories to obtain the Friedmann equations \cite{Saridakis2020,Sheykhi2022,Saridakis2021,Sheykhi2021,Sheykhi2018,Lymperis2018,Sheykhi2023b,Lymperis2021,Almada2022,Almada2022b,Abreu2022,Odintsov2023,Coker2023,Okcu2024b,Ourabah2024}. In particular, the modified Friedmann equations have been investigated in the framework of Barrow entropy\cite{Saridakis2020,Sheykhi2022,Saridakis2021,Sheykhi2021}, Tsallis entropy \cite{Sheykhi2018,Lymperis2018}, and Kaniadakis entropy \cite{Sheykhi2023b,Lymperis2021,Almada2022,Almada2022b,Abreu2022}, and fractional entropy \cite{Coker2023}. Moreover, there has recently been  much attention to investigate the consistency of generalised entropies with observational constraints \cite{Lymperis2018,Almada2022,Almada2022b}.

Another interesting aspect of gravity-thermodynamics conjecture can be regarded as Verlinde's entropic gravity theory \cite{Verlinde2011}. According to this theory, gravity is not a fundamental force, but it is an entropic force whose nature depends on the change in information on the holographic screen. Within this context, Verlinde showed that Newtonian dynamics and gravity can be obtained from entropic gravity. He also showed that Einstein field equation can be derived from entropic gravity. Moreover, entropic gravity theory can be extended to cosmological applications such as the derivation of Friedmann equations \cite{Cai2010}. Studies devoted to entropic gravity have been investigated in the literature \cite{Cai2010,Shu2010,Sheykhi2010c,Sheykhi2011,Gao2010,Ling2010,FCai2010,Basilakos2012,Senay2021,Feng2018,Abreu2018,Jusufi2023}. Beyond the approaches discussed above, Padmanabhan introduced the novel idea that regards the universe as a result of the emergence of cosmic space \cite{Padmanabhan2012}. In his approach, the difference between the degrees of freedom on the boundary and in the bulk is proportional to the change in cosmic volume. Therefore, the space is regarded as an emerging phenomenon as a consequence of the progression of cosmic time. His idea has been extended in various cases \cite{Padmanabhan2012,Cai2012,Yang2012,Sheykhi2013,Eune2013,Dezaki2015}.

It is well-known that other generalised entropy formalisms have also been proposed in the literature \cite{Tsallis2013,Barrow2020,Jalalzadeh2021}. Despite their different setups, these entropies may seem similar. For instance, Tsallis  \cite{Tsallis2013}, Barrow \cite{Barrow2020} and fractional \cite{Jalalzadeh2021} entropies have properties that the entropy may be regarded as the power law function of its area. These similarities suggest that these entropies may belong to a broader family of horizon entropy, that is, they may be obtained from suitable limits of that generalised entropy \cite{Odintsov2023}. Interestingly, the Rényi entropy is similar to EUP modified black hole entropy \cite{Moradpour2019} despite their different origins. Beyond these proposals, recently, a new generalised entropy has been derived from MOND theory by Sheykhi and Liravi \cite{Sheykhi2025}. Starting from a general expression of the MOND theory, the authors of Ref. \cite{Sheykhi2025} obtained a novel entropy from the entropic gravity scenario where the fundamental relation is given by $F\Delta x=T\Delta S$. Then, they obtained the modified Friedmann equations from three different methods, i.e., the first law of thermodynamics at apparent horizon, entropic gravity, and the emergence nature of gravity. In this paper, we refer to this new entropy as the MOND entropy and it is given by \cite{Sheykhi2025}
\begin{equation}
\label{MONDEntropy}
S_{MOND}=\frac{A_{h}}{4}{}_{2}F_{1}\left[\frac{1}{\tilde{\alpha}},\frac{1}{\tilde{\alpha}},\frac{\tilde{\alpha}+1}{\tilde{\alpha}},-\left(\frac{\tilde{\gamma}A_{h}}{4}\right)^{\tilde{\alpha}}\right],
\end{equation}
where $_2F_{1}$ the hypergeometric function, $\tilde{\alpha}$ is a dimensionless parameter and $\tilde{\gamma}$ is a parameter which can be constrained by observations \cite{Sheykhi2025}, and has the dimension of inverse squared length. For $\tilde{\alpha}=1$, this entropy reduces to the well-known Rényi entropy \cite{Moradpour2019}. For $\tilde{\alpha}=2$, this entropy yields a deformed (dual) version of Kaniadakis entropy \cite{Abreu2022}. MOND was originally proposed by Milgrom to explain the rotation curves of spriral galaxies without invoking dark matter \cite{Milgrom1983}. Assuming a new acceleration scale, MOND theory suggests that Newton's second law or the law of gravity deviates from Newtonian case at large distance scales.

Having mentioned that different entropies share common properties and they may be interpreted as different faces of a broader entropy framework, a natural question comes to mind: What is the corresponding modified uncertainty principle associated with this generalised entropy?  Establishing such a connection would enable a unified framework and may provide a new kind of modified uncertainty principle \cite{Du2022}. Since MOND entropy yields Rényi and dual Kaniadakis entropies for the suitable limits, the corresponding modified uncertainty principle must reproduce both entropies in the limiting cases. Moreover, although these entropies do not explicitly imply a cutoff mechanism, they may be endowed with an effective cutoff mechanism due to the corresponding modified uncertainty principle. Therefore, our aim is to obtain a broader modified uncertainty principle which corresponds to MOND entropy. In order to proceed, we first consider the following EUP model \cite{Perivolaropoulos2017}:  
\begin{equation}
\label{HOEUP}
\Delta x\Delta p\geq\frac{1}{2}\frac{1}{1-\alpha\Delta x^{2}},
\end{equation}
where $\alpha$ is a dimensionful EUP parameter and has inverse squared length dimension. Throughout paper, we refer to this EUP as higher-order EUP (HOEUP) since it includes a higher-order correction in position uncertainty compared to simple EUP $\Delta x \Delta p\geq1/2(1+\alpha \Delta x^{2})$. We will derive the Friedmann equations from HOEUP modified entropy-area relation. We will see that these Friedmann equations are the limiting cases of those obtained from MOND entropy \cite{Sheykhi2025}. This indicates that HOEUP can be obtained from MOND entropy. In fact, we will see that HOEUP naturally arises as the perturbative limit of MOND entropy.

The paper is organised as follows: In Section \ref{entropy-area}, we obtain the modified entropy-area relation from HOEUP (\ref{HOEUP}). In Section \ref{firstLawFriedmann}, the modified Friedmann equations are obtained from the first law of thermodynamics at the apparent horizon. Within the framework of entropic gravity, in Section \ref{entropicFriedmann}, we present the modified Friedmann equations. Section \ref{friedmannEmergence} is devoted to deriving the modified Friedmann equations by employing the notion of emergence of cosmic space. In Section \ref{G-second-law}, we check the validity of the GSL. In Section \ref{E-mond}, we establish the connection between MOND entropy and EUP. Starting from MOND entropy (\ref{MONDEntropy}), we derive the corresponding EUP relation. Moreover, we show that this new EUP can yield the EUP models which correspond to Rényi and dual Kaniadakis entropies. Finally, we discuss our results in Section \ref{Conc}. We use the units $\hbar=c=G_{N}=k_{B}=1$ throughout the paper.

\section{Entropy-Area relation from HOEUP} 
\label{entropy-area}

Using the intuitive approach of Ref. \cite{Xiang2009}, we derive the entropy-area relation from HOEUP. Let us begin to consider a particle falling into black hole. Then, the smallest increase in black hole area is characterised by \cite{Bekenstein1973}
\begin{equation}
\label{ChangeOfArea}
\Delta A\sim bm,
\end{equation}
where $m$ and $b$ are the mass and size of the particle, respectively. The above equation can be expressed in a more familiar form, namely the uncertainty principle. To do so, we must consider two limitations on $b$ and $m$. The first limitation is the fact that the width of a wave packet is defined by the particle size, $b\sim \Delta x$. The second limitation is the fact that the momentum uncertainty cannot exceed the particle size, $\Delta p\leq m$. Therefore,  Eq. (\ref{ChangeOfArea}) can be rewritten as
\begin{equation}
\label{ChangeOfArea2}
\Delta A\sim bm\geq \Delta x\Delta p.
\end{equation}
Using the EUP (\ref{HOEUP}) in the above expression yields
\begin{equation}
\label{ChangeOfArea3}
\Delta A_{h}\geq\frac{\gamma}{2\left(1-\alpha\Delta x^{2}\right)},
\end{equation}
where $\gamma$ is proportionality constant which is obtained in the limit $\alpha \rightarrow 0$. Taking $\Delta x\sim 2r_{h}$, EUP modified entropy-area relation is given by
\begin{equation}
\label{EARel}
\frac{dS_{h}}{dA_{h}}\simeq\frac{\Delta S_{min}}{\Delta A_{min}}=\frac{2\ln2\left(1-4\alpha r_{h}^{2}\right)}{\gamma},
\end{equation}
where we use the minimum increase of entropy $\Delta S_{min}=\ln 2$. In the limit $\alpha\rightarrow0$, the above equation must yield the classical case $dS_{h}/dA_{h}=1/4$. Thus, we obtain $\gamma=8\ln 2$. Finally, the EUP modified entropy is obtained from the above equation. It is given by
\begin{equation}
\label{EUPEnt}
S_{EUP}=\pi r_{h}^{2}-2\pi\alpha r_{h}^{4}.
\end{equation}
In the following section, using this entropy, we will obtain the modified Friedmann equations.

\section{Friedmann equations from the first law of thermodynamics}
\label{firstLawFriedmann}

In a compact form, the line element of Friedmann–Lemaitre–Robertson–Walker (FRLW) universe is given by
\begin{equation}
 \label{lineElement}
 ds^{2}=h_{ab}dx^{a}dx^{b}+\widetilde{r}^{2}d\Omega^{2}.
 \end{equation}
Here $\widetilde{r}=a(t)r$, $a(t)$ denotes the scale factor, the coordinates $x^a=(t,r)$, and two-dimensional metric is given by $h_{ab}=diag\left(-1,a^{2}/(1-kr^{2})\right)$. The spatial curvature constant $k=$ $-1$, $0$ and $1$ correspond to  open, flat, and closed universe, respectively. The apparent horizon $\widetilde{r}_{A}$ is defined by \cite{Cai2005}
 \begin{equation}
 \label{apparentHor}
 \widetilde{r}_{A}=\frac{1}{\sqrt{H^{2}+k/a^{2}}},
 \end{equation}
where the Hubble parameter is defined by $H=\dot{a}/a$, and dot denotes the derivative with respect to time. We assume the matter and energy content of the universe as an ideal fluid, which is defined by
 \begin{equation}
 \label{energyMomentumTensor}
 T_{\mu\nu}=(\rho+p)u_{\mu}u_{\nu}+pg_{\mu\nu},
 \end{equation}
where $\rho$, $p$ and $u^{\mu}$ denote energy density, pressure, and four velocity of the fluid. From the conversation of the total energy momentum tensor $\nabla_{\mu}T^{\mu\nu}=0$, the continuity equation is given by
 \begin{equation}
 \label{continuityEqu}
 \dot{\rho}+3H(\rho+p)=0.
 \end{equation}
 Having reviewed the basic properties of FRLW metric, we now turn our attention to the first law of thermodynamics at the apparent horizon. It is given by \cite{Akbar2007}
 \begin{equation}
\label{firstLaw}
dE=TdS+WdV.
\end{equation}
Here, $E=\rho V$ is the total energy, $V$ is the volume surrounded by the horizon, $W$ is the work density. Surface gravity of the apparent horizon is defined by \cite{Cai2005,Hayward1998}
\begin{equation}
\label{kappa}
\kappa =-\frac{1}{\widetilde{r_{A}}}\left (1-
\frac{\dot{\widetilde{r}}_{A}}{2H\widetilde{r}_{A}}\right ).
\end{equation}
Assuming the temperature is proportional to the surface gravity \cite{Akbar2007}, the temperature is given by
\begin{equation}
\label{Temp}
T_{h}=\frac{\kappa}{2\pi}=-\frac{1}{2\pi \widetilde{r}_{A}}\left (1-
\frac{\dot{\widetilde{r}}_{A}}{2H\widetilde{r}_{A}}\right ).
\end{equation} 
Moreover, the work density, which corresponds to work by the volume change of universe, is given by \cite{Hayward1998}
  \begin{equation}
 \label{workDensity}
 W=-\frac{1}{2}T^{ab}h_{ab}=\frac{1}{2}(\rho-p),
 \end{equation}
and volume is given by \cite{Akbar2007}
 \begin{equation}
 V=\frac{4}{3}\pi \widetilde{r}^{3}_{A}.
\label{volume}
\end{equation}
Employing the continuity equation (\ref{continuityEqu}) and volume (\ref{volume}), $dE$ is given by
\begin{equation}
 \label{dE}
 dE=\rho dV+Vd\rho=4\pi\rho\widetilde{r}^{2}_{A}d\widetilde{r}_{A}-4\pi(\rho+p)\widetilde{r}^{3}_{A}Hdt.
 \end{equation}
From Eqs. (\ref{workDensity}) and (\ref{volume}), we can obtain the $WdV$ term as follows:
 \begin{equation}
\label{WdV}
WdV=2\pi(\rho-p)\widetilde{r}^{2}_{A}d\widetilde{r}_{A}.
\end{equation}
Since entropy is a function of area $S=S(A_{h})$, the general expression of entropy is expressed by \cite{Awad2014}
\begin{equation}
\label{f(A)Rel}
S_{h}=\frac{f(A_{h})}{4}.
\end{equation}
Then, the differential form of entropy is given by
\begin{equation}
\label{diffOfEntropy}
\frac{dS_{h}}{dA_{h}}=\frac{f'(A_{h})}{4},
\end{equation}
where prime is the derivative with respect to $A_{h}$. Comparing Eq. (\ref{EARel}) with Eq. (\ref{diffOfEntropy}), we obtain
\begin{equation}
\label{f'A}
f'(A_{h})=1-4\alpha\widetilde{r}^{2}_{A}.
\end{equation}
Thus, using Eqs. (\ref{Temp}), (\ref{diffOfEntropy}) and (\ref{f'A}), the term $T_{h}dS{_h}$ is given by
\begin{equation}
\label{TdS}
T_{h}dS_{h}=-\left(1-\frac{\dot{\widetilde{r}}_{A}}{2H\widetilde{r}_{A}}\right)(1-4\alpha\widetilde{r}^{2}_{A})d\widetilde{r}_{A}.
\end{equation}
By inserting Eqs. (\ref{dE}), (\ref{WdV}) and (\ref{TdS}) into the first law of thermodynamics at the apparent horizon (\ref{firstLaw}) and using the differential form of the apparent horizon 
\begin{equation}
\label{diffRA}
d\widetilde{r}_{A}=-H\widetilde{r}^{3}_{A}\left(\dot{H}-\frac{k}{a^{2}}\right)dt
\end{equation}
one can obtain the following dynamical equation
\begin{equation}
\label{dynamicalEq}
4\pi(\rho+p)\widetilde{r}^{3}_{A}Hdt=f'(A_{h})d\widetilde{r}_{A}.
\end{equation}
Merging the continuity equation (\ref{continuityEqu}) with the above equation yields the differential form of the first Friedmann equation
\begin{equation}
 \label{firstEquation}
\frac{f'(A_{h})}{\widetilde{r}^{3}_{A}}d\widetilde{r}_{A}=-\frac{4\pi}{3}d\rho.
\end{equation}
Integrating this equation, we get the first Friedmann equation as 
\begin{equation}
\label{firstFriedmann}
\frac{1}{2\widetilde{r}_{A}^{2}}+4\alpha\ln\left(2\sqrt{\alpha}\widetilde{r}_{A}\right)=\frac{4\pi\rho}{3},
\end{equation}
where we set the integration constant to $-4\alpha\ln\left(2\sqrt{\alpha}\right)$, so that the argument of logarithm is dimensionless, since $\boldsymbol{\alpha}$ has the dimension of inverse squared length. At last, the second Friedmann equation can be obtained from Eqs. (\ref{f'A}), (\ref{diffRA}) and (\ref{dynamicalEq}). It is given by
\begin{equation}
\label{secondFriedmann}
\left(1-4\alpha\widetilde{r}_{A}^{2}\right)\left(\dot{H}-\frac{k}{a^{2}}\right)=-4\pi\left(\rho+p\right).
\end{equation}
Using the definition of the apparent horizon (\ref{apparentHor}), the first Friedmann equation can be expressed in terms of the Hubble parameter as follows:
\begin{equation}
\label{firstFriedmannH}
\left(H^{2}+\frac{k}{a^{2}}\right)-4\alpha\ln\left[\frac{1}{4\alpha}\left(H^{2}+\frac{k}{a^{2}}\right)\right]=\frac{8\pi\rho}{3}.
\end{equation}
This is the Renyi modified Friedmann equation \cite{Sheykhi2025}. For $\alpha=\pi\tilde{\gamma}/4$, this equation exactly reproduces  the Friedmann equations obtained from MOND entropy at  $\tilde{\alpha}=1$.

\section{Friedmann equations from the entropic gravity}
\label{entropicFriedmann}

Following the arguments of Refs. \cite{Verlinde2011,Cai2010}, we are going to obtain the modified Friedmann equations within the framework of the entropic gravity scenario. Let us consider a spatial region $\mathcal{V}$ enclosed by a holographic screen on the boundary $\mathcal{\partial V}$. The number of bits on the holographic screen is identified as \cite{Verlinde2011} 
\begin{equation}
\label{NARel}
N=A_{h}.
\end{equation}
From $S_{h}=A_{h}/4$ and the above equation, the relation between the entropy and the number of bits can be expressed as
\begin{equation}
N=4S_{h}.
\label{NSRelation}
\end{equation}
Furthermore, we suppose that the total energy of the holographic screen is defined by the equipartition law of energy
\begin{equation}
\label{equPart}
E=\frac{1}{2}NT,
\end{equation}
where  the screen temperature $T$ corresponds to Unruh temperature. It is given by
\begin{equation}
\label{unTemp}
T=\frac{a_{r}}{2\pi}=-\frac{\ddot{a}r}{2\pi}.
\end{equation}
 Here, $a_{r}$ is the acceleration, and is given by\cite{Cai2010}
\begin{equation}
\label{accl}
a_{r}=-\frac{d^{2}\widetilde{r}_{A}}{dt^{2}}=-\ddot{a}r.
\end{equation}
In order to obtain the Friedmann equation in general relativity, we must consider the active gravitational mass inside the horizon \cite{Cai2010}
\begin{equation}
\label{KomarMss}
\mathcal{M}=2\int_{\mathcal{V}}dV\left(T_{\mu\nu}-\frac{1}{2}Tg_{\mu\nu}\right)u^{\mu}u^{\nu}=\frac{4}{3}\pi(\rho+3p)\widetilde{r}^{3}_{A}.
\end{equation}
where the energy-momentum tensor is given by Eq. (\ref{energyMomentumTensor}). From EUP-modified entropy (\ref{EUPEnt}) and Eq. (\ref{NSRelation}), the modified number of bits is given by
\begin{equation}
\label{modifiedN}
N=4\pi\tilde{r}_{A}^{2}-8\pi\alpha\tilde{r}_{A}^{4}.
\end{equation}
Under assumption $\mathcal{M} = E$ and employing Eqs. (\ref{equPart})-(\ref{modifiedN}), we obtain the acceleration equation:
\begin{equation} 
\label{e-adot/a-2}
\frac{\ddot{a}}{a} = \frac{-4 \pi}{3} ( \rho + 3 p) 
\frac{1}{1-2 \alpha \widetilde{r}^{2}_{A}}.
\end{equation}
Multiplying $2\dot{a}a$ on both sides of the second Friedmann equation and using the continuity equation, we get
\begin{equation}
\label{integForm}
\int\frac{d\dot{a}^{2}}{dt}dt=\frac{8\pi}{3}\int\frac{1}{1-2\alpha a^{2}r^{2}}\frac{d\left(\rho a^{2}\right)}{dt}dt.
\end{equation}
In order to proceed  and solve this integral, we choose the equation of state as $p=\omega \rho$. Employing the equation of state in continuity equation (\ref{continuityEqu}), we obtain
\begin{equation}
\label{solutionContiEq}
\rho=\rho_{0}a^{-3(1+\omega)},
\end{equation}
where $\rho_{0}$ is a constant. From the above equation, we find
\begin{equation}
\label{dRhoa2}
d(\rho a^{2})=-\rho_{0}(1+3\omega)a^{-2-3\omega}da.
\end{equation}
Employing this equation in the integral, we obtain
\begin{equation}
\label{integForm2}
\int\frac{d\dot{a}^{2}}{dt}dt=-\frac{8\pi(1+3\omega)\rho_{0}}{3}\int\frac{a^{-3\omega-2}}{1-2\alpha a^{2}r^{2}}\frac{da}{dt}dt.
\end{equation}
Finally, by solving this integral, we obtain the first Friedmann equation as follows:
\begin{equation}
\label{firstFriedmannEG}
\left(\frac{\dot{a}}{a}\right)^{2}+\frac{k}{a^{2}}=\frac{8\pi\rho}{3}{}_{2}F_{1}\left[1,-\frac{1+3\omega}{2},\frac{1-3\omega}{2},2\alpha\widetilde{r}_{A}^{2}\right],
\end{equation}
where ${}_{2}F_{1}$ is the hypergeometric function. For $\alpha=-\pi\tilde{\gamma}/2$, this equation exactly coincides with the Friedmann equations obtained from MOND entropy at $\tilde{\alpha}=1$ \cite{Sheykhi2025}.

\section{Friedmann equations from emergence of cosmic space} 
\label{friedmannEmergence}

Here, we are going to obtain the modified Friedmann equation by followig the arguments of Padmanabhan \cite{Padmanabhan2012}. In his approach, $N_{sur}=N_{bulk}$ is valid for a pure de Sitter universe, where $N_{sur}$ and $N_{bulk}$ denote degrees of freedom on the boundary and in the bulk, respectively. For an asymptotically de Sitter universe, he proposed that the difference between $N_{sur}$ and $N_{bulk}$ is proportional to the increase of cosmic volume $dV$. For an infinitesimal interval of cosmic time, the increase of cosmic volume is defined by
\begin{equation}
\label{dVRelation}
\frac{dV}{dt}=\left(N_{sur}-N_{bulk}\right).
\end{equation}
Using $T=\frac{H}{2\pi}$ and $V=\frac{4\pi}{3H^{3}}$, Padmanabhan obtained the Friedmann equation for flat case. Later, the generalisation to the non-flat cases, proposed by Sheykhi \cite{Sheykhi2013}, is given by
\begin{equation}
\label{dVRelationRevised}
\frac{dV}{dt}=\widetilde{r}_{A}H\left(N_{sur}-N_{bulk}\right).
\end{equation}
For a flat universe, this equation reduces to Eq. (\ref{dVRelation}) since  $\tilde{r}_{A}H=1$. The temperature of the surface is given by \cite{Cai2005}
\begin{equation}
\label{tempAprrox}
T_{h}=\frac{1}{2\pi\widetilde{r}_{A}}.
\end{equation}
Here the temperature is just an approximation of Eq. (\ref{Temp}) since the volume change $dV$ is associated with the infinitesimal time interval $dt$. From the entropy (\ref{EUPEnt}), the EUP-modified number of degrees of freedom on the surface is given by
\begin{equation}
\label{NSur}
N_{sur}=4S_{h}=4(\pi \widetilde{r}^{2}_{A}-2\pi\alpha \widetilde{r}^{4}_{A}).
\end{equation}
Total energy enclosed by apparent horizon is the Komar energy and is defined by
\begin{equation}
\label{KomarEnergy2}
E_{Komar}=|(\rho +3p)|V,
\end{equation}
where the volume is given by Eq. (\ref{volume}). The bulk degrees of freedom are defined by the equipartition law of energy. It is given by
\begin{equation}
\label{Nbulk2}
N_{bulk}=\frac{2|E_{Komar}|}{T}.
\end{equation}
For an accelerated universe, we consider $\rho+3p<0$. Combining the above equation with the Komar energy, we obtain $N_{bulk}$ as
\begin{equation}
\label{Nbulk}
N_{bulk}=-\frac{16\pi^{2}\widetilde{r}^{4}_{A}}{3}(\rho+3p).
\end{equation}
Employing $N_{sur}$ and $N_{bulk}$ in Eq. (\ref{dVRelationRevised}), we obtain the following expression
\begin{equation}
\label{intStepEMS}
\frac{2\tilde{r}_{A}^{-3}\dot{\tilde{r_{A}}}}{H}=2\tilde{r}_{A}^{-2}-\alpha+\frac{8\pi}{3}\left(\rho+3p\right).
\end{equation}
Multiplying  both sides by $\dot{a}a$ and using the continuity equation, Eq. (\ref{intStepEMS}) can be expressed as
\begin{equation}
\label{intStepEMS2}
-\frac{d\left(\tilde{r}_{A}^{-2}a^{2}\right)}{dt}+\alpha a\frac{da}{dt}=-\frac{8\pi}{3}\frac{d\left(\rho a^{2}\right)}{dt}
\end{equation}
After integration, the above expression leads to the first Friedmann equation in terms of the apparent horizon:
\begin{equation}
\label{FriedmannRA}
-\frac{1}{\tilde{r}_{A}^{2}}+\frac{\alpha}{2}=-\frac{8\pi\rho}{3}.
\end{equation}
Using Eq. (\ref{apparentHor}), we obtain
\begin{equation}
\label{FriedEqCosSp}
\left(H^{2}+\frac{k}{a^{2}}\right)-\frac{\alpha}{2}=\frac{8\pi\rho}{3}.
\end{equation}
For $\alpha=\pi \tilde{\gamma}$, this equation exactly yields the Friedmann equation derived from MOND entropy at $\tilde{\alpha}=1$ \cite{Sheykhi2025}.

\section{Generalised second law of thermodynamics} 
\label{G-second-law}

Let us check the validity of GSL for the EUP-modified entropy. GSL states that the sum of horizon and matter fields entropies cannot decrease with time. Eq. (\ref{dynamicalEq}) can be written 
\begin{equation}
	\label{GSL1}
	\dot{\widetilde{r}}_{A}=\frac{4\pi(\rho+p)H\widetilde{r}_{A}^{3}}{1-4\alpha\widetilde{r}^{2}_{A}}.
\end{equation}
Using this equation and Eq. (\ref{f'A}) in Eq. (\ref{TdS}), we find
\begin{equation}
\label{GSL2}
T_{h}\dot{S_{h}}=4\pi\left(\rho+p\right)H\tilde{r}_{A}^{3}\left(1-\frac{2\pi(\rho+p)\tilde{r}_{A}^{2}}{1-4\alpha\tilde{r}_{A}^{2}}\right)
\end{equation}
If only the entropy of the horizon is taken into consideration, the second law may be violated for the accelerated expansion $\rho+p<0$. Thus, we must consider the entropy of matter fields inside the horizon. To do so, we begin with the Gibbs equation. It is given by \cite{Izquierdo2006}
\begin{equation}
\label{GibbsEquation}
T_{m}dS_{m}=d(\rho V)+pdV=Vd\rho+(\rho+p)dV,
\end{equation}
where the subscript  $m$ denotes the matter fields. We assume the thermal equilibrium between the horizon and the matter fields $(T_{h}=T_{m})$. Therefore, there is no energy flow between the horizon and matter fields. Combining Eqs. (\ref{volume}), (\ref{continuityEqu}), (\ref{GSL1}) and (\ref{GibbsEquation}), we find
\begin{equation}
\label{GSL3}
T_{h}\dot{S_{m}}=-4\pi\left(\rho+p\right)H\tilde{r}_{A}^{3}\left(1-\frac{4\pi\left(\rho+p\right)\tilde{r}_{A}^{2}}{1-4\alpha\tilde{r}_{A}^{2}}\right).
\end{equation}
From Eqs. (\ref{GSL2}) and (\ref{GSL3}), the total entropy is given by
\begin{equation}
\label{GSL4}
T_{h}\left(\dot{S_{h}}+\dot{S_{m}}\right)=\frac{8\pi^{2}\left(\rho+p\right)^{2}H\tilde{r}_{A}^{5}}{1-4\alpha\tilde{r}_{A}^{2}}.
\end{equation}
The GSL holds for condition $0<\tilde{r}_{A}<\frac{1}{2\sqrt{\alpha}}$. This condition ensures the positivity of the denominator $1-4\alpha\tilde{r}_{A}^{2}$. The upper bound naturally arises from EUP. In fact, HOEUP gives the maximum position uncertainty at $\Delta x=\frac{1}{\sqrt{3\alpha}}.$ Therefore, the maximum apparent horizon is given by $\tilde{r}_{A}^{max}=\frac{1}{2\sqrt{3\alpha}}$. Consequently, the GSL is always satisfied within the allowed range of the apparent horizon. At this point, we give some comments on the maximum position uncertainty. In Ref. \cite{Perivolaropoulos2017}, the author interpreted the horizontal asymptote of HOEUP as the maximum position uncertainty. In our analysis, we do not adopt this interpretation. Instead, we define the maximum position uncertainty at $\Delta x=\frac{1}{\sqrt{3\alpha}}$ where the minimum momentum uncertainty occurs. Just like other EUP models, the reason for this choice is that the definition of IR cutoff depends on minimum momentum uncertainty or equivalently corresponding maximum position uncertainty. Thus, our adoption may ensure a coherent interpretation across different models. This point will be more clear in the next section.

\section{EUP from MOND} \label{E-mond}

In the previous sections, we have obtained the modified Friedmann equations through three different approaches: the first law of thermodynamics at the apparent horizon \cite{Akbar2007}, entropic gravity case \cite{Verlinde2011,Cai2010}  and the emergent cosmic space framework \cite{Padmanabhan2012,Sheykhi2013}. Interestingly, for $\tilde{\alpha}=1$, these equations exhibit a close resemblance to MOND entropy modified Friedmann equations in Ref. \cite{Sheykhi2025}. It is worth noting that the case $\tilde{\alpha}=1$ corresponds to Rényi entropy modified Friedmann equations \cite{Sheykhi2025}. Motivated by this correlation, in reverse engineering style, starting from the  MOND entropy, we would like to obtain the modified uncertainty principle. This approach not only establishes the direct link between EUP and MOND, but also exhibits how large scale correction to uncertainty principle can emerge from a generalised entropy framework. Let us begin the differential form of MOND entropy (\ref{MONDEntropy}), which is given by
\begin{equation}
\label{MondEARel}
\frac{dS_{h}}{dA_{h}}\simeq\frac{\Delta S_{min}}{\Delta A_{min}}=\frac{1}{4}\left(1+\left(\frac{\tilde{\gamma}A_{h}}{4}\right)^{\tilde{\alpha}}\right)^{-1/\tilde{a}}.
\end{equation}
The minimum change in the horizon area is given by \cite{Bekenstein1973,Xiang2009}
\begin{equation}
\label{AminMond}
\Delta A_{min}=4\ln 2\left(1+\left(\frac{\tilde{\gamma}A_{h}}{4}\right)^{\tilde{\alpha}}\right)^{1/\tilde{\alpha}}=4\ln 2\left(1+\left(\frac{\tilde{\gamma}\pi\Delta x^{2}}{4}\right)^{\tilde{\alpha}}\right)^{1/\tilde{\alpha}},
\end{equation}
Here, we use $A_{h}=4\pi r_{h}^{2}$ with $r_{h}=\Delta x/2$, and minimum change in entropy is $\Delta S_{min}=\ln 2$. Recalling Eq. (\ref{ChangeOfArea2}), $\Delta A_{min} \simeq\gamma\Delta x \Delta p$, one can write the following expression:
\begin{equation}
\label{EUPMONDCons}
4\ln 2\left(1+\left(\frac{\tilde{\gamma}\pi\Delta x^{2}}{4}\right)^{\tilde{\alpha}}\right)^{1/\tilde{\alpha}}\simeq\gamma\Delta x \Delta p,
\end{equation}
where we take $\gamma=8\ln2$. From this relation, the lower bound of momentum uncertainty $\Delta p_{0}$ is expressed by
\begin{equation}
\label{EUPMONDCons2}
\Delta p\geq\Delta p_{0}\simeq\frac{1}{2\Delta x}\left[1+\left(\frac{\tilde{\gamma}\pi\Delta x^{2}}{4}\right)^{\tilde{\alpha}}\right]^{1/\tilde{\alpha}}
\end{equation}
Multiplying both sides of this inequality by $\Delta x$, we arrive at the final form of EUP
\begin{equation}
\label{EUPMONDFINAL}
\Delta x\Delta p\geq\frac{1}{2}\left[1+\left(\frac{\tilde{\gamma}\pi\Delta x^{2}}{4}\right)^{\tilde{\alpha}}\right]^{1/\tilde{\alpha}},
\end{equation}
which is a new EUP derived from MOND entropy (\ref{MONDEntropy}). We refer to this new EUP as MOND EUP. The MOND entropy (\ref{MONDEntropy}) reduces to the Rényi entropy for $\tilde{\alpha}=1$ and the dual Kaniadakis entropy for $\tilde{\alpha}=2$ \cite{Sheykhi2025}. Consistently, the MOND entropy also reproduces the EUP forms that are related to the Rényi and dual Kaniadakis entropies. For $\tilde{\alpha}=1$, the MOND EUP reduces to the simplest EUP \cite{Moradpour2019}
\begin{equation}
\label{EUPSimple}
\Delta x\Delta p\geq\frac{1}{2}\left(1+\frac{\tilde{\gamma}\pi\Delta x^{2}}{4}\right),
\end{equation}
which yields the Rényi entropy \cite{Moradpour2019}
\begin{equation}
\label{RényiEntropy}
S_{h}=\frac{1}{\tilde{\gamma}}\ln\left(1+\frac{\tilde{\gamma} A_{h}}{4}\right).
\end{equation}
Solving Eq. (\ref{EUPSimple}) for $\Delta x$ yields two solutions,
\begin{equation}
\label{EUPSols}
\Delta x=\frac{4\Delta p\pm2\sqrt{4\Delta p^{2}-\pi\tilde{\gamma}}}{\pi\tilde{\gamma}}.
\end{equation}
For the limit $\tilde{\gamma}\rightarrow 0$, the solution with the minus sign in front of the square root reduces to the standard uncertainty principle, while the other solution diverges.

\begin{figure}
	\centerline{\includegraphics[width=10cm]{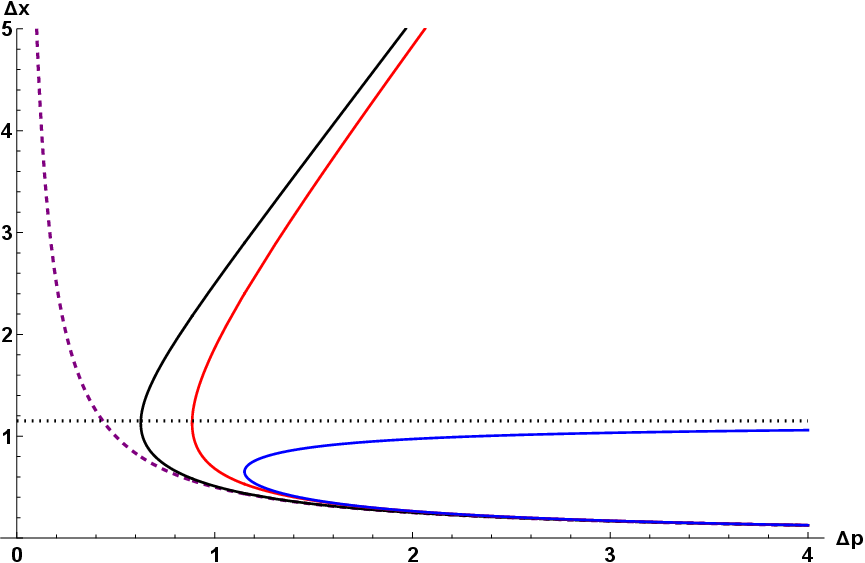}}
	\caption{$\Delta x$ versus $\Delta p$ for different EUPs. The blue, red, and black curves correspond to HOEUP, EUP and DKEUP, respectively, while the dashed-purple curve correspond to standard uncertainty principle. The horizontal asymptote is represented by dotted-black line. We set $\tilde{\gamma}=1$ and use the units $\hbar=c=G_{N}=k_{B}=1$.} 
	\label{EUPS}
\end{figure}
For $\tilde{\alpha}=2$, the MOND EUP yields the new kind of EUP form
\begin{equation}
\label{DualKaniadakisEUP}
\Delta x\Delta p\geq\sqrt{1+\left(\frac{\tilde{\gamma}\pi\Delta x^{2}}{4}\right)^{2}},
\end{equation}
which reproduces the dual Kaniadakis entropy \cite{Abreu2022}
\begin{equation}
\label{KaniadakisEntropy}
S_{h}=\frac{1}{\tilde{\gamma}}\ln\left(\frac{\tilde{\gamma}A_{h}}{4}+\sqrt{1+\left(\frac{\tilde{\gamma} A_{h}}{4}\right)^{2}}\right).
\end{equation}
Therefore, we may refer this EUP as dual Kaniadakis EUP (DKEUP). This new EUP leads to a quartic equation in $\Delta x$, which yields four solutions. However, two solutions are unphysical since they are negative. The remaining two solutions are given by
\begin{equation}
\label{DKEUPSols}
\Delta x=\frac{2\sqrt{8\Delta p^{2}\pm\sqrt{64\Delta p^{4}-\pi^{2}\tilde{\gamma}^{2}}}}{\pi\tilde{\gamma}}.
\end{equation}
Similarly, smaller solution reduces to the standard uncertainty principle while larger solution diverges for $\tilde{\gamma}\rightarrow0$ . In addition, HOEUP (\ref{HOEUP}) can be derived from MOND entropy.  Expanding MOND entropy (\ref{MONDEntropy}), we find \cite{Sheykhi2025}
\begin{equation}
\label{EUPEntropyMOND}
S_{h}=\frac{A_{h}}{4}\left\{ 1-\frac{\tilde{\gamma}^{\tilde{\alpha}}}{\tilde{\alpha}(\tilde{\alpha}+1)}\left(\frac{A_{h}}{4}\right)^{\tilde{\alpha}}+...\right\} 
\end{equation}
where $\frac{\tilde{\gamma}^{\tilde{\alpha}}}{\tilde{\alpha}(\tilde{\alpha}+1)}<<1$. Here we use the series expansion of hypergeometric function given by \cite{Sheykhi2025}
\begin{equation}
{}_2F_1\left[a,b;c,z\right] = \sum_{n=0}^{\infty}
\frac{(a)_n (b)_n}{(c)_n\,n!}\,z^n = 1 + \frac{ab}{c}\,z + \mathcal{O}(z^2).
\end{equation}
For $\tilde{\alpha}=1$ and $A_{h}=4\pi r_{h}^{2}$, we find 
\begin{equation}
\label{HOEUPEntropyEUP}
S_{h}=\pi r_{h}^{2}-\frac{\tilde{\gamma}\pi^{2}r_{h}^{2}}{2}. 
\end{equation}
Following the same steps as in the previous derivation, the EUP corresponding to this entropy is given by
\begin{equation}
\label{HOEUPEUP}
\Delta x\Delta p\geq\frac{1}{2}\frac{1}{1-\frac{\tilde{\gamma}\pi(\Delta x)^{2}}{4}}.
\end{equation}
For $\alpha=\frac{\tilde{\gamma}\pi}{4}$, this entropy and its EUP exactly correspond to Eqs. (\ref{EUPEnt}) and (\ref{HOEUP}), respectively. Thus, we can obtain HOEUP as the perturbative limit of MOND entropy.

In Fig. \ref{EUPS}, we present different uncertainty principles in $\Delta x-\Delta p$ plane. In the case of standard uncertainty principle (dashed-purple curve), $\Delta p$ is allowed to approach zero, which results in $\Delta x\rightarrow\infty$. It is clear that the standard uncertainty does not imply a minimum momentum uncertainty. In contrast, EUP models imply a minimum momentum uncertainty. As can be seen in Fig. \ref{EUPS}, all EUPs have minimum momentum uncertainty. Moreover, all EUPs restrict the allowed region in the phase space. The size of the allowed region is the most restricted for the HOEUP model (blue curve). The EUP (red curve) and DKEUP (black curve) models also exhibit bounded regions, but the sizes of the allowed regions are larger than in the HOEUP model. The reason why the most restricted case occurs in HOEUP model is the presence of a horizontal asymptote at $\Delta x=2/\sqrt{\pi\tilde{\gamma}}$. The upper branch of HOEUP cannot exceed the horizontal asymptote (dotted-black line), thus leading to the most restricted region. This unique property is a direct consequence of the functional structure of HOEUP model. Unlike HOEUP, the other EUP models do not exhibit such a behaviour, and  the upper branches grow with increasing $\Delta p$. In Table \ref{tab:EUP-forms}, we present the all EUP models and their maximum position uncertainty $(\Delta x)_{max}$ and minimum momentum uncertainty $(\Delta p)_{min}$. The values of $\Delta x_{max}$ are obtained from the lower branches of the solutions. Simple EUP and DKEUP yield the same maximum position uncertainty, while HOEUP model gives a smaller value. HOEUP yields the largest $(\Delta p)_{min}$, while DKEUP yields the smallest $(\Delta p)_{min}$.
\begin{table}[ht!]
\centering
\setlength{\tabcolsep}{6pt}
\renewcommand{\arraystretch}{1.25}
\small
\begin{tabular}{C{1.8cm} C{7.5cm} C{2.6cm} C{2.8cm}}
\toprule
\textbf{Name} & \textbf{Form} & \multicolumn{1}{c}{\boldmath$(\Delta x)_{\max}$} & \multicolumn{1}{c}{\boldmath$(\Delta p)_{\min}$} \\
\midrule
EUP \cite{Moradpour2019}&
\( \Delta x\,\Delta p \ge \tfrac{1}{2}\!\left[ 1 + \tfrac{\tilde{\gamma}\pi}{4}\,(\Delta x)^{2} \right] \) &
\( \tfrac{2}{\sqrt{\tilde{\gamma}\pi}} \) &
\( \tfrac{\sqrt{\tilde{\gamma}\pi}}{2} \) \\
DKEUP &
\( \Delta x\,\Delta p \ge \tfrac{1}{2}\,\sqrt{ 1 + \left( \tfrac{\tilde{\gamma}\pi}{4}\,(\Delta x)^{2} \right)^{2} } \) &
\( \tfrac{2}{\sqrt{\tilde{\gamma}\pi}} \) &
\( \tfrac{1}{2}\sqrt{\tfrac{\tilde{\gamma}\pi}{2}} \) \\
HOEUP \cite{Perivolaropoulos2017} &
\( \Delta x\,\Delta p \ge \tfrac{1}{2}\,\tfrac{1}{1 - \frac{\tilde{\gamma}\,\pi\,(\Delta x)^{2}}{4}} \) &
\( \tfrac{2}{\sqrt{3\tilde{\gamma}\pi}} \) &
\( \tfrac{3\sqrt{3\tilde{\gamma}\pi}}{8}  \) \\
\bottomrule
\end{tabular}
\caption{Different forms of EUP and their corresponding \((\Delta x)_{\max}\) and \((\Delta p)_{\min}\) values. We use the units }$\hbar=c=G_{N}=k_{B}=1$.
\label{tab:EUP-forms}
\end{table}

Note that both Rényi and dual Kaniadakis entropies do not imply the existence of an IR cutoff by themselves. However, once they are associated with their corresponding EUP models, these generalised entropies attain such a feature. In this case, the corresponding EUPs introduce a maximum position uncertainty $(\Delta x)_{max}$ and a minimum momentum uncertainty $(\Delta p)_{min}$, thus endowing the entropies with an IR cutoff. Consequently, one may regard the IR cutoff as an effective feature of entropies, which is inherited through their underlying EUP frameworks.

Before ending this section, we give some comments on the parameter $\tilde{\gamma}$, which is given by $\tilde{\gamma}=\frac{a_{0}}{\pi G M}$ \cite{Sheykhi2025}. Here,  $a_{0}$ is MOND acceleration scale, typically of order $a_{0} \sim 10^{-10}\,\mathrm{m\,s^{-2}}$ \cite{Sheykhi2025}, while $M$ is the characteristic mass associated with the system under consideration. For example, $M$ can be identified with the baryonic mass of a galaxy or can be interpreted as an effective mass enclosed by the cosmological horizon. In this sense, the value of $\tilde{\gamma}$ depends on the physical scale of the system. The latter case suggests that $\tilde{\gamma}$ introduces a large scale momentum cutoff in the extended uncertainty principle. Thus, MOND may be interpreted  as a natural IR cutoff at cosmological scale.

\section{Conclusions}
\label{Conc}

In this study, we have revealed the deep connections between EUP models and recently proposed MOND entropy \cite{Sheykhi2025}.  Beginning from HOEUP \cite{Perivolaropoulos2017}, we derived the modified entropy-area relation, which forms the basis for subsequent derivations of Friedmann equations within the framework of gravity-thermodynamics conjecture: first, by applying the first law of thermodynamics at the apparent horizon; second, within the entropic gravity framework; and third, through the emergent cosmic space perspective. We have further investigated the validity of GSL, showing that it holds within the allowed region of the apparent horizon. Interestingly, the Friedmann equations in this paper correspond to the limiting cases ($\tilde{\alpha}=1$) of those obtained from the MOND entropy \cite{Sheykhi2025}. This correspondence clearly implies that HOEUP can be regarded as the limiting case of MOND entropy. Therefore, starting from MOND entropy, we have obtained the corresponding extended uncertainty principle.   

Remarkably, this new novel EUP consistently reproduces the EUPs associated with Rényi and dual Kaniadakis entropies in the limiting cases. It also naturally yields HOEUP in the perturbative limit. Moreover, although Rényi and dual Kaniadakis entropies do not explicitly imply IR cutoff, they can still have IR cutoff as an effective feature due to their corresponding EUP. This result implies that different generalised entropies may gain new properties that are evident only at a deeper level, that is, the underlying modified uncertainty principle. In this sense, the transition from MOND entropy to corresponding EUP provides an unifying framework that bridges different entropies, IR cutoff mechanisms, and cosmological dynamics in a consistent manner.   

Overall, this work provides a systematic framework for investigating a generalised entropy and its underlying modified uncertainty principle in a unified manner. As discussed in the Introduction, different entropies may be regarded as various faces of a broader entropy formalism. This becomes even more clear at the deeper level when searching the corresponding modified uncertainty principle as a complementary framework. Therefore, our study may shed light on the unifying role that connects the generalised entropies with their corresponding modified uncertainty principles.

\section*{Acknowledgments}

E. A. is grateful to Professor Paul J. Steinhardt and Princeton University for warm hospitality. E. A. also acknowledges T\"{U}B\.{I}TAK since this study is partially supported by T\"{U}B\.{I}TAK under 2219 Project: "\textit{Studies in the framework of the Chaotic Cyclic Cosmology}". The authors are grateful to the anonymous reviewers for their valuable comments and constructive suggestions.

%\section*{Data Availability}
%No data was used for the research described in the article.

%\section*{Conflict of interest }
%The authors declare that they have no conflict of interest.

\end{document}